\documentclass[a4paper,10pt]{article}
\usepackage[margin=1in]{geometry}
\usepackage[T1]{fontenc}
\usepackage[utf8]{inputenc}
\usepackage{authblk}
\usepackage[square]{natbib}
\usepackage{graphicx}

\title{Global Climate Modeling of the Martian water cycle with improved microphysics and radiatively active water ice clouds}

\author[1]{Navarro T.}
\author[1,2]{Madeleine J.-B.}
\author[1]{Forget F.}
\author[1]{Spiga A.}
\author[1]{Millour E.}
\author[3]{Montmessin F.}
\author[3]{M\"a\"att\"anen A.}
\affil[1]{Laboratoire de M\'et\'eorologie Dynamique (LMD), CNRS/UPMC/IPSL, Paris, France.}
\affil[2]{Previously at Department of Geological Sciences, Brown University, Providence, Rhode Island, USA.}
\affil[2]{Laboratoire Atmosph\`eres, Milieux, Observations Spatiales (LATMOS), CNRS/UVSQ/IPSL, Guyancourt, France.}

\date{}

\begin{document}

\maketitle

\begin{abstract}


Water ice clouds play a key role in the radiative transfer of the Martian atmosphere, impacting its thermal structure, its circulation and, in turn, the water cycle.
Recent studies including the radiative effects of clouds in Global Climate Models (GCMs) have found that the corresponding feedbacks amplify the model defaults.
In particular it prevents model with simple microphysics to reproduce even the basic characteristic of the water cycle. Within that context,
we propose a new implementation of the water cycle in GCMs, including a detailed cloud microphysics taking into account nucleation on dust particles, ice particle growth,
and scavenging of dust particles due to the condensation of ice. 
We implement these new methods in the Laboratoire de M\'et\'eorologie Dynamique GCM and find satisfying agreement with the Thermal Emission Spectrometer observations of both water vapor and
cloud opacities, with a significant improvement when compared to GCMs taking into account radiative effects of water ice clouds without this implementation.
However, a lack of water vapor in the tropics after Ls=180$^\circ$ is persistent in simulations compared to observations, as a consequence of aphelion cloud radiative effects strengthening
the Hadley cell.
Our improvements also allow us to explore questions raised by recent observations of the Martian atmosphere.
Supersaturation above the hygropause is predicted in line with SPICAM (SPectroscopy for Investigation of Characteristics of the Atmosphere of Mars) observations.
The model also suggests for the first time that the scavenging of dust by water ice clouds alone fails to fully account for the detached dust layers observed by the Mars Climate Sounder.

\end{abstract}

\section{Introduction}

The Martian water cycle is an important component of the Martian climate. The repeatable seasonal cycle of water vapor was first characterized
from orbit by the Viking Mars Atmospheric Water Detector (MAWD) \citep{Jako:82}, and later monitored for several years by the Thermal Emission Spectrometer (TES) onboard Mars Global Surveyor (MGS) \citep{Smit:01tes,Smit:04}, 
as well as the Mars Express instruments \citep{Fedo:06,Fouc:07,Tsch:08,Malt:11}.
TES was also used to monitor the water ice cloud opacities around 2 pm local time.
In light of these measurements, a first generation of numerical models able to simulate the main characteristics 
of the observed cycle
was developed by introducing water vapor sublimation, condensation and transport in three dimensional Global
Climate Model (GCM). \cite{Rich:02water} first described the major mechanisms controlling the cycle.
\cite{Mont:04} obtained similar results and analyzed the importance of clouds in the overall water transport.
However, these models did not take into account the radiative effect of water ice clouds which had long been assumed to 
negligible \citep{Zure:92}.

It is only recently that clouds have been acknowledged to play a key role on the temperature structure of the Martian atmosphere. \citet{Habe:99}, \citet{Cola:99} and \citet{Hins:04clouds} showed that they could create thermal inversions and explain temperature profiles observed in locations where clouds were present. \citet{Wils:08} showed that the atmospheric temperatures in the equatorial region could only be fully explained by taking into account the radiative effect of clouds.
Finally \citet{Made:12radclouds} studied the local and global effects of radiatively active clouds in the LMD GCM and showed that they help reduce global temperature biases between model and observations throughout northern spring and summer.
This was also briefly discussed by \citet{Urat:13h2o} using a new GCM derived from the National Center for Atmospheric Research (NCAR) Community Atmosphere Model.
However, \citet{Made:12radclouds}, \citet{Urat:13h2o}, as well as unpublished studies performed with other Martian climate models (NASA Ames GCM: \citealt{Habe:11}; Geophysical Fluid Dynamic Laboratory (GFDL) GCM: R. J. Wilson, personal communication) found that taking into account the radiative effects of clouds had a strong negative impact on the modeled water cycle which cannot be simulated realistically.
In most cases, it tended to be much drier than without radiative effects of clouds.
\citet{Habe:11} attributed this to unrealistic spurious water ice clouds that form above the Northern polar cap during summer, affect the global radiation budget at the surface, and therefore reduce the total amount of water vapor released in the atmosphere.
\citet{Urat:13h2o} also found that ``it was extremely difficult to simultaneously duplicate the observed cloud opacities in the high summer latitudes due to the highly coupled nature of the polar cap temperature, cloud formation, atmospheric heating and cooling by clouds, and the saturation vapor pressure''.
They concluded that their model failed ``to take into account some process that occurs at high latitudes in the summer hemisphere that suppresses cloud formation'', and suggested that this process could be supersaturation.
They were inspired by the observations of \citet{Malt:11} who reported the presence of water vapor in supersaturation above the hygropause using Mars Express SPICAM (SPectroscopy for Investigation of Characteristics of the Atmosphere of Mars) solar occultation. In most GCM, including the LMD GCM until today,
such supersaturations were not taken into account.
Condensation was assumed to occur when saturation is reached, and all the excess water converted into ice.



In this paper we investigate the effects of radiatively active clouds on the atmospheric water ice and vapor in a GCM that consistently model the nucleation of cloud particles and their growth in supersaturated atmosphere.
We show that it prevents the formation of water ice clouds above the Northern polar cap during summer.
A description of the new water cycle model (including the detailed microphysics scheme, as well as a better representation of permanent ice caps) is given in section \ref{secdescr}
Section \ref{secresults} describes the GCM results, the necessity for the microphysical modeling of clouds, and the influence of tunable parameters.
Section \ref{secdiscuss} compares simulations to TES measurements and discusses the various interactions and features of the water cycle.

\section{Model Description} \label{secdescr}

The model used in this paper is the Mars General Circulation Model (GCM) developed at Laboratoire de M\'et\'eorologie Dynamique (LMD) \citep{Forg:99,Forg:11}

The dynamical core that integrates the fluid hydrodynamic equations for the atmosphere is a finite difference numerical model.
The horizontal discretization is a longitude-latitude grid and the vertical discretization uses hydrid $\sigma$-pressure coordinates. 
Due to the longitude-latitude grid, the Courant-Friedrichs-Lewy criterion is violated in upper
latitudes, and a classical Fourier filter is applied at the pole to the dynamical variables,
such as the horizontal wind, the temperature and the surface pressure \citep{Forg:99}
\\
The physical parametrization of the model takes into account emission and absorption of CO$_2$ gas in infrared, as well as emission, absorption and scattering of two aerosols: airborne dust \citep{Made:11} and water ice \citep{Made:12radclouds}.
The seasonal variations of atmospheric mass by condensation of CO$_2$ \citep{Forg:98} reproduces the pressure cycle as measured by Viking lander \citep{Hess:80}.
A subgrid scale convection using a thermal plume model is parametrized in the planetary boundary layer \citep{Cola:13}. 
A photochemical module with 15 species \citep{Lefe:04,Lefe:08} is implemented in the model, and is used only in one specific case in section \ref{supersat}.
The model has an extension to the thermosphere up to the exobase around 250~km
\citep{Ange:05,Gonz:09a}, but it is not activated in the simulations of the present paper.
\\
The ground is modeled with 18 subsurface layers down to 18 meters with a horizontally and vertically varying thermal inertia to solve the thermal conduction for underground thermal inertia values ranging from 30 to 2000 Jm$^{-2}$K$^{-1}$s$^{-1/2}$.
The CO$_2$ cycle is tuned thanks to, among other things, the CO$_2$ ice albedo and the underground thermal inertia in the model, representing the underground water ice centimeters below the surface from mid-latitude to the pole in both hemispheres \citep{Habe:08pss}.
\\
Airborne dust distribution is simulated using a ``semi-interactive'' scheme \citep{Made:11}:
the shape of the dust vertical profiles freely evolves with atmospheric mixing, but the total column values are rescaled to match the observed opacities as compiled by \cite{Mont:14}.
The transported dust is modeled using a two-moment scheme \citep{Made:11}.
The two moments are the number mixing ratio dust particles (number of particles per kg of air) and the mass mixing ratio (kg per kg of air).
Each quantity is advected independently and, assuming a lognormal size distribution, the effective radius of dust particles can be computed.
A constant and homogeneous dust lifting value is used as a source of dust, balanced by a size-dependant gravitational sedimentation. Below, we describe how this scheme is now 
used to parametrize the scavenging of dust particles by water ice condensation.

\subsection{Ground Water Ice}

\subsubsection{Permanent Ice Reservoirs}
The model uses thermal inertia and albedo derived from TES measurements \citep{Putz:07}. 
However, the permanent water ice reservoirs on the northern polar ice cap, the main source of Martian atmospheric water in the model, 
are modeled with their own prescribed albedo and thermal inertia, 
instead of using surface values derived from spacecraft data. 
Using true ice thermal inertia and albedo data are indeed more suitable to a GCM, 
as it simulates actual perennial surface water ice, instead of relying on measurements that mix bare ground and ice within a GCM grid mesh.

Albedo and thermal inertia values of this ice are left as tunable parameters because one cannot exactly know the ice surface purity, although
acceptable values range from 0.3 to 0.5 for albedo \citep{Wils:07} and 500 to 2000 J.m$^{-2}$.K$^{-1}$.s$^{-1/2}$ \citep{Putz:07} for thermal inertia.

The extent of a permanent ice reservoir in the GCM is simply the area of the corresponding mesh defined by the longitudinal-latitudinal grid.
One of the main parameters for water vapor release is the surface temperature controlled by direct solar illumination, or equivalently the latitude of the permanent ice reservoir, which can be used to define permanent reservoirs.

\subsubsection{Amount of permanent ice}\label{permice1}
\begin{figure}
\centering
\noindent\includegraphics[width=40pc]{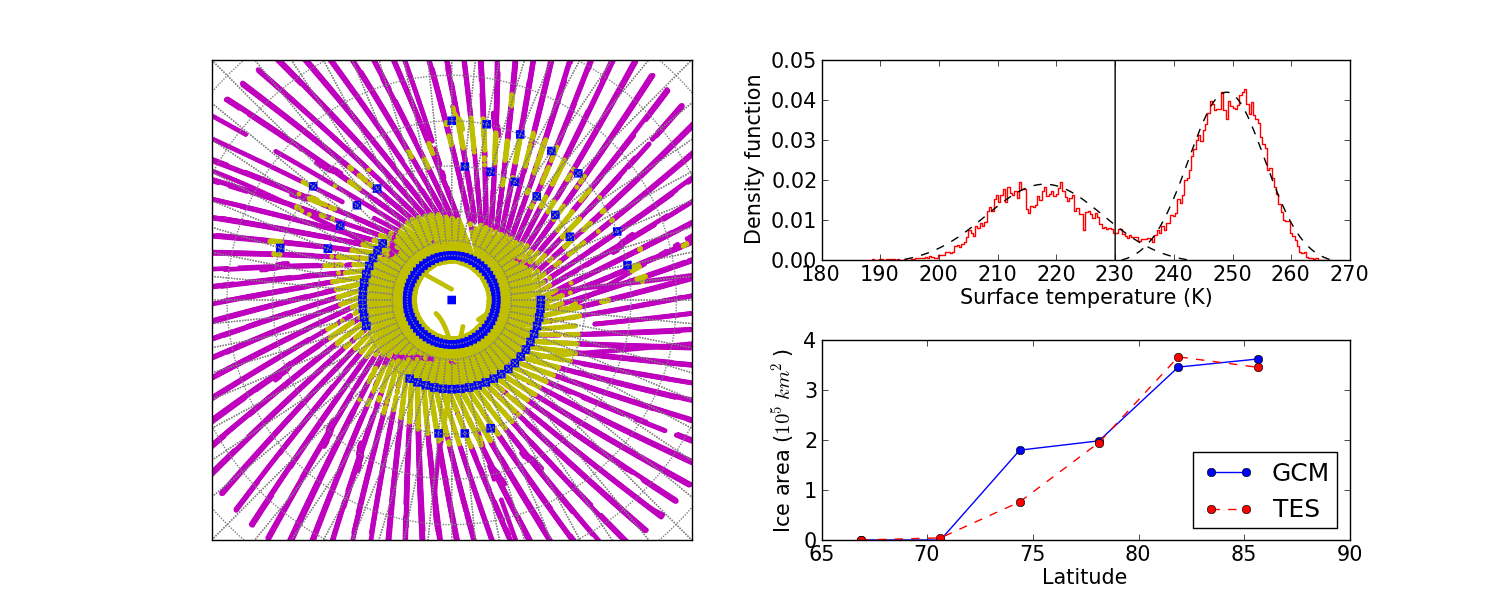}
\caption{
Surface temperatures between Ls=90$^\circ$ and Ls=120$^\circ$ for latitudes above 65$^\circ$N and at local times between 10 am and 5 pm.
Geographic distribution of daytime temperatures from TES between Ls=110$^\circ$ and Ls=115$^\circ$ above 65$^\circ$N colder than 230 K (beige) and warmer (pink) and representation in the GCM (left panel).
Distribution of these temperatures (red) and two gaussian fits (dashed) of the underlying cold and warm distributions (upper right panel).
Ice as a function of latitude from TES temperatures and as prescribed in the GCM (lower right panel).}
\label{figD}
\end{figure}

A new representation of perennial surface ice is used to better constraints on the water cycle than the simple 
distributions usually used in GCMs \citep{Rich:02water,Mont:04}.
The actual locations occupied by perennial surface water ice can be derived from summertime 
TES surface temperature because water ice tends to remain significantly colder than bare ground
during northern summer as illustrated on the upper right panel of figure \ref{figD}.

In practice, we assumed that permanent ice is present when surface temperatures remains below 230 K above 
65$^\circ$N between Ls=90$^\circ$ and Ls=120$^\circ$, and that the corresponding GCM grid points would be covered by water ice.
This results in a realistic spatial distribution of permanent ice in the GCM with a good approximation of the ice variation with latitude
The lower right panel of Figure \ref{figD} compares the total area of ice at a given latitude in the GCM (in the resolution used in this paper) and 
as retrieved from the observations (corrected to account for the imperfect spatial coverage). The modeled ice coverage is in good agreement with the observations,
except at latitude 75$^\circ$N, where the ice is overestimated in the GCM. However, as explained in section~\ref{spaceparams} this overestimation does not results 
in an overestimation of water vapour release. 

\subsubsection{Seasonal Frost} \label{frost}
As for polar ice reservoirs, seasonal frost can modify surface albedo.
We use a threshold value of 5 microns, as in \citet{Rich:02water} and \citet{Mont:04}, of surface ice deposition to change surface albedo from bare soil albedo to the same value prescribed for
the perennial north polar cap reservoirs.
This choice was made in order to reduce the number of tunable parameters.
In reality, both albedos are significantly variable in space and time (in the future it may be possible to improve - and complicate - the model by better representing the perennial surface ice and the frost albedo).
The maximal extent of the observed seasonal frost to latitudes around 50$^\circ$N and 50$^\circ$S is reproduced.
The diffusion of water vapor into the regolith is not included.

\subsubsection{Sublimation Scheme} \label{sublimationscheme}
The sublimation of surface ice deposits is controlled by the surface turbulent flux \citep{Forg:99,Mont:04}:
\begin{equation}
\label{flasar}
E_w=\rho C_d u_*(q_{sat}-q_{w})
\end{equation}
where $E_w$ is the turbulent flux of water into the lowest atmospheric model layer, $\rho$ is the atmospheric density near the ground, $C_d$ is the aerodynamic drag coefficient, $u_*$ is the friction velocity, and $q_{w}$ is the mass mixing ratio of water vapor ($q_{sat}$ is the saturation mass mixing ratio at surface temperature) in the first layer.
$C_d$ depends, among other variables, on the surface roughness coefficient and on the Richardson number (see \citealt{Cola:13}).
Equation \ref{flasar} shows that the amount of water ice that is sublimated depends on both the humidity and the intensity of the turbulent mixing close to the surface. 

\subsection{Microphysics}

\subsubsection{Cloud Microphysics under Martian Conditions} \label{intromicrophy} 
In the previous studies performed with the GCM \citep{Mont:04,Made:12radclouds}, the cloud scheme was rather simplistic: water vapor is turned instantaneously into ice, or the other way around, to reach the saturation pressure and the number of ice particles is defined as a fixed fraction of available dust particles.
However, it is well known that in the atmosphere the water vapor phase does not suddenly turn into a condensed phase as soon as saturation, mainly controlled by temperature, is reached.
Instead, ice particles nucleate and grow onto airborne dust \citep[See e.g.][]{Maat:05}. 
Such processes have a tight control of the properties of the clouds as well as on the remaining vapor phase.
Nucleation eventually determines the number of dust nuclei onto which water vapor will condense.
Inclusion of these two processes makes predictions for cloud particle size physically consistent and also allows the existence of substantial amounts of supersaturation (e.g. when dust nuclei are too sparse).
We included a microphysics scheme adapted from \citet{Mont:02} in the GCM. 
The main motivations are to address the effect of dust scavenging by water ice clouds, allow supersaturation of water vapor, and correctly predict water ice clouds characteristics.

\subsubsection{Nucleation}
As on Earth, airborne dust is expected to play the role of condensation nuclei under Martian atmospheric conditions \citep{Good:86}, referred below as \emph{CCN} for Cloud Condensation Nuclei.
Therefore, we assume that nucleation occurs within the frame of the theory of heterogeneous nucleation.
As described in \citet{Mont:02}, the number of activated CCN by nucleation depends on a poorly constrained contact parameter m, given by $m = \cos \theta$ with $\theta$ the ``contact angle''
(in liquid phase it is the angle of the interface between the droplet and its nucleus). 
In the case of a nucleation as efficient as possible, the contact would be maximized with $m$ equal to 1.
In reality, this contact parameter depends on the intrinsic properties of water and Martian dust. 
Laboratory experiments on Mars dust analogs have typically found values ranging between 0.97 and 0.93, or even lower values (down to $m=0.84$) 
at temperatures below 180~K, with nucleation becoming more difficult when lowering temperatures \citep{Trai:09,Irac:10,Pheb:11}. 
The sensitivity of the modeled water cycle with $m$ is discussed in section~\ref{params}.
\\
In practice, in the GCM, the dust particles which are activated are transferred into a new kind of tracers, the CCN.
The quantity and size distribution of CCN is described in each model boxes by their own mass mixing ratio and number mixing ratio 
assuming a lognormal distribution and an effective variance of 0.1 
(in other words we use the two-moment scheme of \citet{Made:11} for the dry dust particles, and add the CCN size distribution that differs from the dust size distribution).
To implement the nucleation process itself (i.e. to compute the mass and the number of dust particles which become CCN at every timestep in a model gridbox) 
we first discretize the dust distribution around the mean radius.
Then, in each radius bin, we perform nucleation as described in section~2.1 in \cite{Mont:02}.
The new distributions of dust and CCN is the sum over the radius bins for the mass and the number of particles.
We use 5 bins for this discretization from 0.1 to 50 microns. Tests did not show a significative difference when using 50 bins.

\subsubsection{Condensation} \label{sectioncondensation}
We use the formulation of \citet{Mont:02} for the growth of ice crystals: 
\begin{equation}
\frac{dr}{dt} =\frac{1}{r}\left( \frac{S-S^{*}}{R_{c}+R_{d}} \right) \label{drdt}
\end{equation}
with $r$ the radius of the crystal (assumed to be spherical), $S$ the saturation ratio between water vapor partial pressure and the saturation pressure, $S^{*}$ the saturation ratio at equilibrium taking into account the crystal surface curvature (Kelvin effect) and $R_{c}$ and $R_{d}$ the heat and diffusive resistances described in \citet{Mont:04} and \citet{Mont:02}.

We initially introduced equation \ref{drdt} and the formulation for nucleation of \citet{Mont:02} in the GCM using a 15 minutes timestep with an explicit numerical scheme.
This led to a severe underestimation of $r$ for ice clouds formed in the model and an overestimation of the amount of CCN involved.
Underestimating the radius has dramatic effects on the cloud evolution in case of radiatively active clouds.
Indeed, if the same amount of ice is condensed on a greater number of particles, the resulting cloud has smaller particles, and hence higher opacity.
In particular, any local drop in temperature due to the radiative effect of clouds is overestimated, leading to increased saturation ratio (even if the amount of water vapor decreases), reinforcing nuclei activation.
This positive feedback loop between cloud formation and temperature, controlled by the ice particles size and number, yields unrealistic cloud opacities.
Again, radiatively active clouds do not only modify Martian dynamics on a large scale \citep{Hins:04clouds,Made:12radclouds} but also the structure and the formation of the clouds itself due to microphysical processes on the local scale.
Therefore, it appears that the use of an elaborate microphysical scheme in a GCM needs to be followed by a careful understanding of the ice particle size dependence to the integration timestep.
Various tests with single column model and 3-D simulations led us to conclude that it is necessary to couple both nucleation and condensation through sub-timesteps no larger than one minute.
In particular, this was necessary to prevent the formation of unrealistic clouds above the North pole during summer, as explained in section \ref{globalmicrophy}.

\subsubsection{Scavenging}
Scavenging is a key physical process on the Earth, allowing to ``clean'' the atmosphere from aerosols through water precipitation.
On Mars, scavenging is thought to play a role in the dust cycle \citep{Clan:96b}, but detailed studies are lacking. 
Modeling scavenging of dust particles working as cloud nuclei is straightforward in our new cloud microphysical scheme is implemented: CCN are advected like any other tracer in the GCM. 
As nucleation occurs, dust particles are converted to CCN, incorporated into ice crystals, while the number of remaining ``ice-free'' dust particles is reduced accordingly. 
CCN sedimentate along with the ice particles and are released only after all of the ice crystals have sublimated, returning to the ``ice-free'' dust inventory.

In our model, one specificity is that 
the mass and number of dust particles are multiplied by a conversion factor at each timestep so that the dust column matches the observed column-opacities (see \citealt{Made:11}).
We use the same conversion factor for CCN when it comes to getting actual values of mass and number at each timestep.
This implies two assumptions: 

1. CCN do not affect radiative transfer. In other words, dust captive into ice crystals does not interact with light and extinction by ice dominates. 

2. The typical lifetime of CCN in ice particles is lower than the typical time associated with major dust changes.
Indeed the conversion factor used for dust varies with time and space according to changes in prescribed dust opacity.
If a dust particle is sequestrated in an ice crystal, the conversion factor is then based on the local dust conditions, and not on the conditions where and when the dust particle was captured by ice.

Within and below the clouds, the scavenging of dust by falling crystals which could intercept and coalesce with the dust particles is neglected. 
The mass flux associated with this process can be estimated following \citet{Sein:86}.
On Mars, it is 2 to 3 orders of magnitude smaller than the sedimentation flux of CCN.

\section{Global Water Cycle Sensitivity to Model Parameters} \label{secresults}

\subsection{Parameters} \label{params}
Modeling the water cycle requires the setting of values for four poorly constrained model parameters: the albedo of surface ice and frost, the thermal inertia of 
perennial water ice deposits, the contact parameter $m$ and the
variance of the size distribution of the ice cloud particles which affects the radiative properties of the clouds and the sedimentation flux of ice cloud particles. 
This last point has a strong impact on the model results because the cloud particle size effective variance $\nu_{eff}$ 
is directly used to calculate the sedimentation flux via an ``effective'' fall radius $r^{sed}=r_c(1+\nu_{eff})^3$ used in the Stokes-Cunningham relationship for falling particles \citep{Mont:04}, with r$_c$ the mass mean radius
This reflects the fact that, when it comes to the mass transport of a population of particles due to sedimentation, r$_c$ is biased toward larger particles.

The three reference diagnostics, collected from observations, for the tuning values of the four model parameters are 1) the sublimation peak of water vapor 
during northern summer solstice of 60 pr.$\mu$m, 2) the zonal mean of the amount of atmospheric water vapor in the tropics during northern fall of 10 pr.$\mu$m, 
and 3) the mean aphelion cloud belt opacity of 0.15 at 825 cm$^{-1}$ at a local time of 2 pm.
These diagnostics have been chosen because they refer to quantities that are handy for physical interpretation and they well constrain the simulated water cycle.
The sublimation peak controls most of the total atmospheric water vapor throughout the year on the whole planet, the opacity of the aphelion cloud belt constrains the impact of clouds on atmospheric temperatures \citep{Made:12radclouds}, and the tropical water vapor is a good indicative of meridional transport of vapor from northern polar regions.

\subsection{Reference Run} \label{refrun}
The results in this section are obtained by running the LMD GCM at a resolution of $64 \times 48$ on the longitude-latitude grid and 29 levels reaching a maximum altitude above ground of 0.02 Pa, about 80 km. 
The column dust opacity is prescribed at a given location and time using a scenario derived from observational data \citep{Mont:14}.
The scenario for Martian Year (MY) 26 has been used by default, because it is representative of years without a global dust storm. 
Results for other years are addressed in section \ref{globalmicrophy}.

The tunable parameters have been set to the values compiled in table \ref{refrunparams}.
These specific values were chosen in order to obtain a realistic water cycle. 
The overall simulated water cycle is compared to TES observations \citep{Smit:04} during MY26.

\begin{table}
\centering
\begin{tabular}{|c|c|}
\hline
Parameter & value \\
\hline
Contact parameter $m$ & 0.95 \\
\hline
Effective variance for sedimentation $\nu_{eff}$ & 0.1 \\
\hline
Albedo of frost and permanent ice & 0.35 \\
\hline
Thermal inertia of permanent ice & 800 tiu \\
\hline
\end{tabular}
\caption{Parameters used for the reference run}
\label{refrunparams}
\end{table}

\begin{figure}
\centering
\noindent\includegraphics[width=40pc]{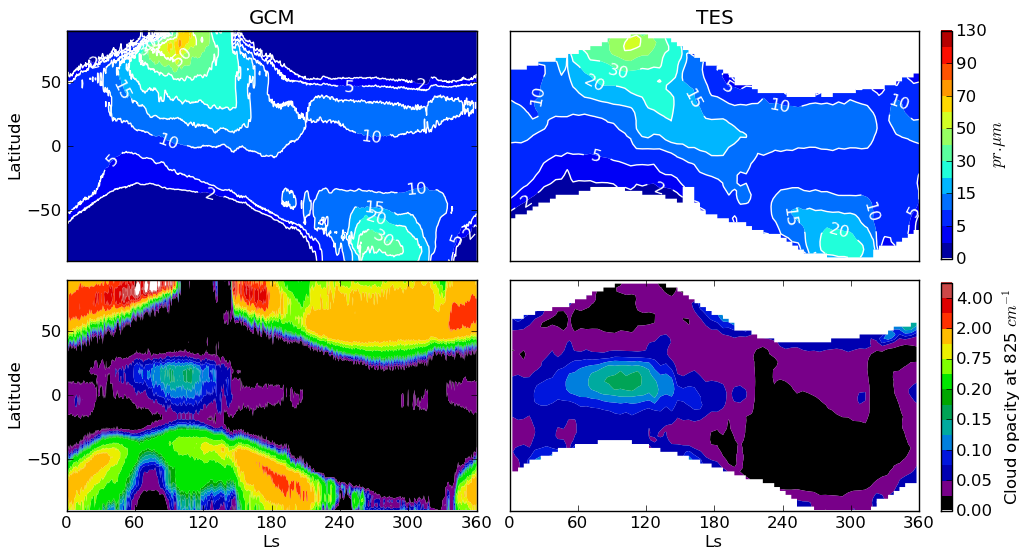}
\caption{Zonal mean quantities at 14h local time for MY26 of atmospheric water vapor as simulated by the GCM (top left) and observed by TES (top right), GCM cloud opacity (bottom left) and TES cloud opacity (bottom right)}
\label{figA}
\end{figure}

An equilibrated state for the water cycle is reached after 20 years of simulation (Figure \ref{figA}).
The main difference between the modeled zonal average water vapor fields and the TES observations is the summer sublimation peak at the North pole, whose amplitude is 30\% too large, 
and the amount of water vapor in the Northern hemisphere tropics after Ls=180$^\circ$, which is too low in the GCM.
The difference in timing for the peak of the frost sublimation in the Southern hemisphere is mainly controlled by the seasonal timing of the CO$_2$ ice cap sublimation. 

\subsection{Global effect of the improved microphysics}\label{globalmicrophy}

\begin{figure}
\centering
\noindent\includegraphics[width=40pc]{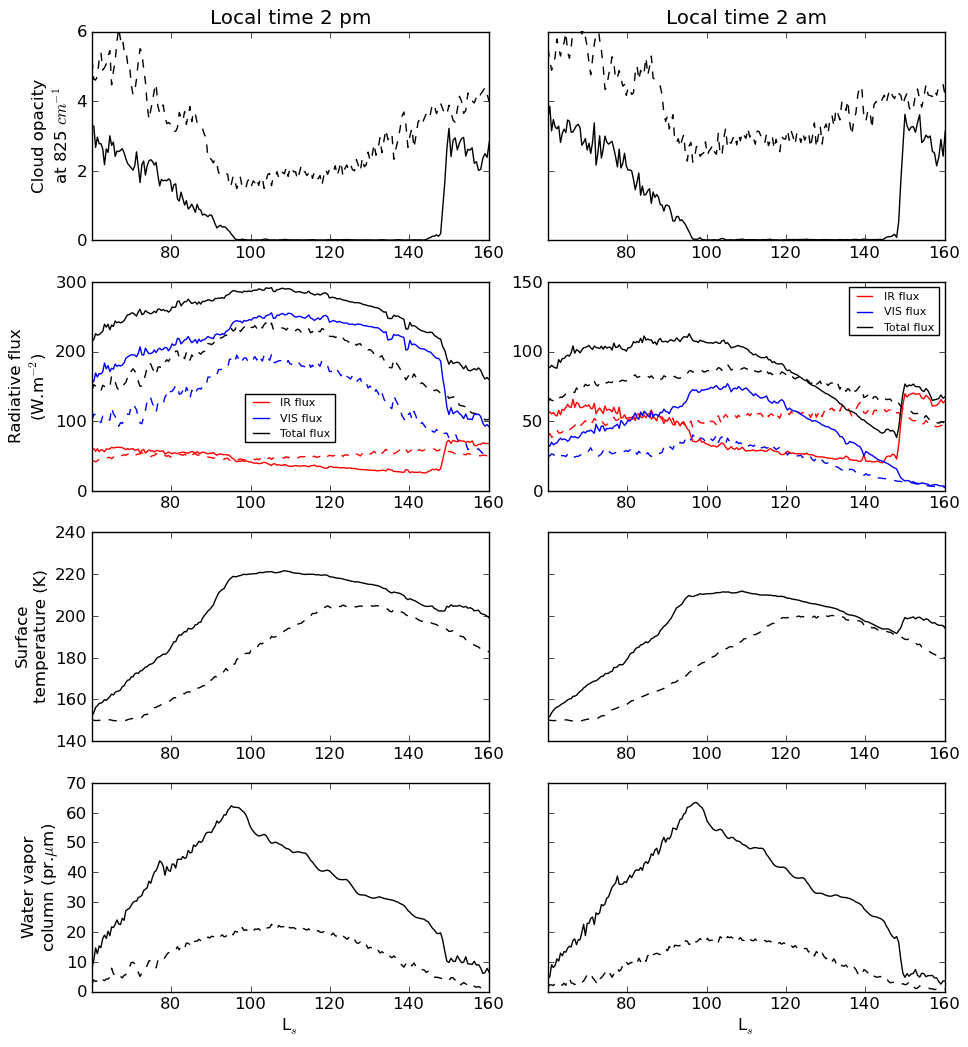}
\caption{Zonal mean quantities for latitudes 75$^\circ$N to 90$^\circ$N for simulations with microphysics (plain) and without (dashed), between Ls=60$^\circ$ and Ls=160$^\circ$. Left column shows values at local time 2 pm and right column at local time 2 am. The first row shows the cloud opacity at 825 cm$^{-1}$. The second row shows the total radiative flux received at the surface (black), and its infrared (red) and visible (blue) contributions. The third row shows the surface temperatures for the perennial surface ice only.
The fourth row shows the atmospheric water vapor. Note that the vertical axis is not the same for the radiative flux budget at the two different local hours, unlike the other rows.
}
\label{figZ}
\end{figure}

The effect of the microphysics on the water cycle of the reference run is fundamental.
This is illustrated in figure \ref{figZ} for two different simulations.

The first one is the reference run, and the second one is the same simulation, except that the cloud microphysics scheme is not used and supersaturated water vapor is turned instantaneously into
ice with a predefined number of ice particles, as said in \cite{Mont:04}.
The cloud opacity shows that, for this second simulation, the polar hood clouds persist throughout the summer.
As mentioned above, this situation is not realistic, as observations by the MGS Mars Orbiter Camera \citep{Wang:02} or TES \citep{Smit:02} show that clouds above the Northern polar cap during summer are intermittent, scattered across the polar region and very faint.
For the reference run, we see a quasi-absence of polar clouds between Ls=100$^\circ$ and Ls=145$^\circ$, as expected from observations.
When comparing the two simulations, the infrared downward radiative flux is greater with the simulation without microphysics, while the visible flux is smaller.
This is simply due to the fact that clouds limit the visible sunlight that reach the surface and they emit in the infrared band.
The total downward radiative flux to the surface is substantially smaller for the simulation without microphysics, except at local time 2 am after Ls = 120$^\circ$.
This explains why the surface of perennial surface ice is colder for the simulation without microphysics, at all local times during the whole summer.\\
As a consequence, the surface ice is sublimated in smaller quantities in the simulation without microphysics.
Therefore, the column-integrated water vapor in the atmosphere is subsequently decreased for the simulation without microphysics.
When the polar clouds re-appear after Ls=145$^\circ$ in the reference simulation, there is a decrease of the atmospheric water vapor because it is trapped in the clouds.
Polar clouds also increase the total 2 am radiative flux and the surface temperature after Ls=145$^\circ$, but the surface ice sublimation is not enhanced enough to overcompensate the loss of water vapor.
\\
In this example, the continuous presence of polar clouds in the simulation without microphysics causes a decrease of the amount of water vapor released in the atmosphere,
as in \cite{Habe:11}, \cite{Made:12radclouds} and in the GFDL GCM (R. J. Wilson, personnal communication).
This behaviour is very dependent on the properties of clouds simulated in the model, and could also enhance the amount of water vapor if the total flux happens to be increased when compared to the reference simulation, if the infrared radiative flux to the surface caused by clouds overcompensates the loss of visible flux.
The only realistic way to prevent such feedback loops is by having clear-sky conditions during the summer thanks to the use of the microphysics, such as the reference simulation illustrates it.
\\
One should also be aware that the simulation without microphysics induces a global water cycle that is not in a steady state, and the next years of simulation would have less and less water vapor at the pole, and on the whole planet.

It is worth noting that even in the reference simulation, polar clouds that exist before Ls=100$^\circ$ seem to still play a role on the surface temperature and the amount of atmospheric water vapor.
There is an extreme sensitivity of the whole water cycle to the critical short period of the summer solstice at the North pole as already noted by \citet{Urat:13h2o}.

\subsection{Sensitivity to Parameters} \label{sensitivityparams} 

\subsubsection{Surface ice albedo and thermal inertia}
\begin{figure}[!ht]
\centering
\noindent\includegraphics[width=40pc]{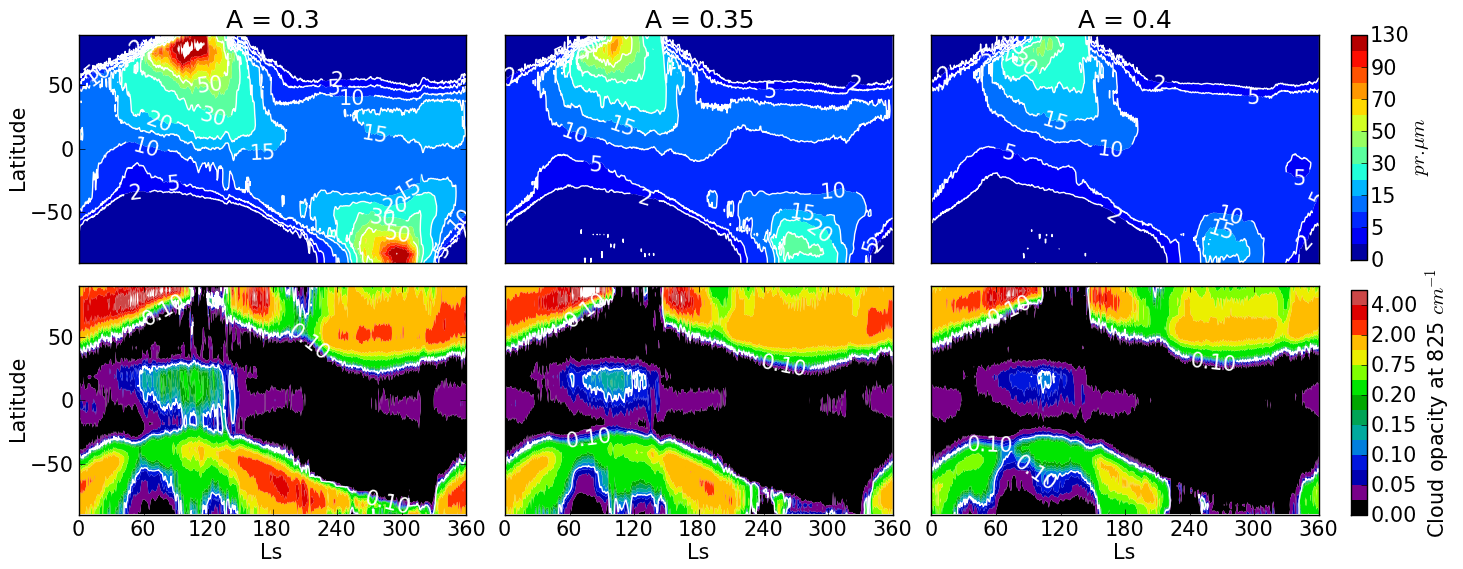}
\caption{GCM zonal mean quantities during daytime for MY26 of atmospheric water vapor (top row) and cloud opacity (bottom row) with three different values of albedo for perennial surface ice : 0.3 (left), 0.35 (middle) and 0.4 (right)}
\label{figP2}
\end{figure}

\begin{figure}[!ht]
\centering
\noindent\includegraphics[width=40pc]{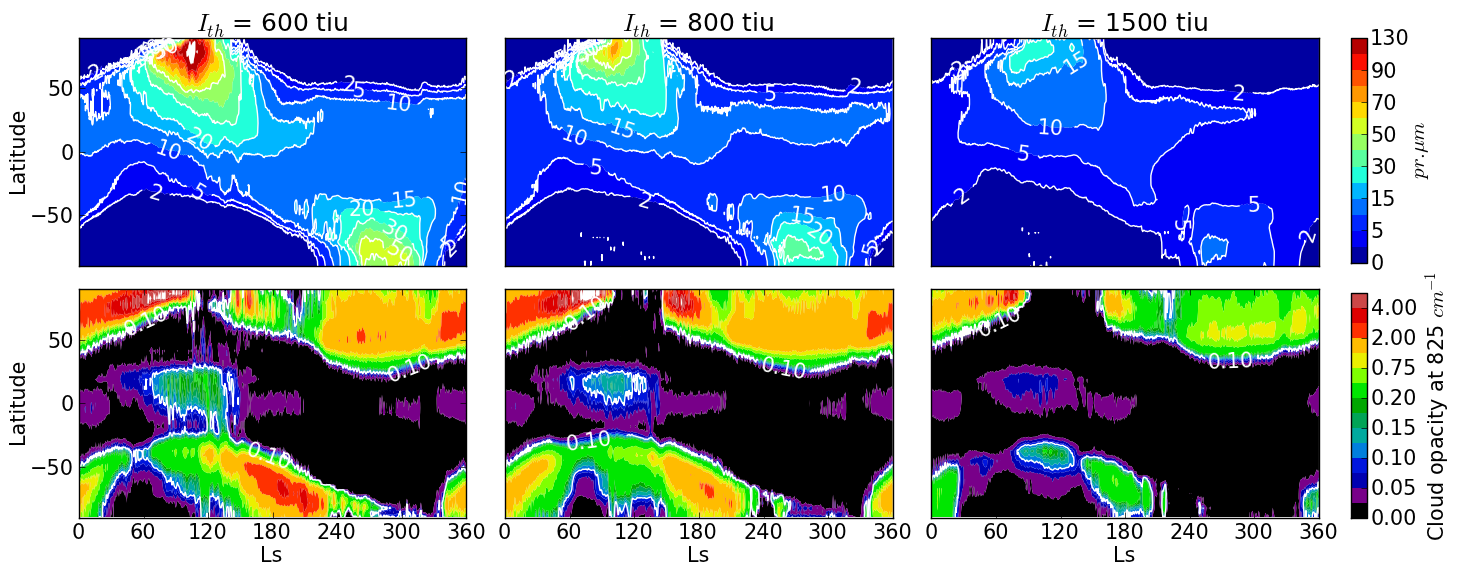}
\caption{GCM zonal mean quantities during daytime for MY26 of atmospheric water vapor (top row) and cloud opacity (bottom row) with three different values of thermal inertia for perennial surface ice : 600 t.i.u. (left), 800 t.i.u. (middle) and 1500 t.i.u. (right).
N.B.: :w
tiu = J.m$^{-2}$.K$^{-1}$.s$^{-\frac{1}{2}}$}
\label{figP4}
\end{figure}

\begin{figure}[!ht]
\centering
\noindent\includegraphics[width=40pc]{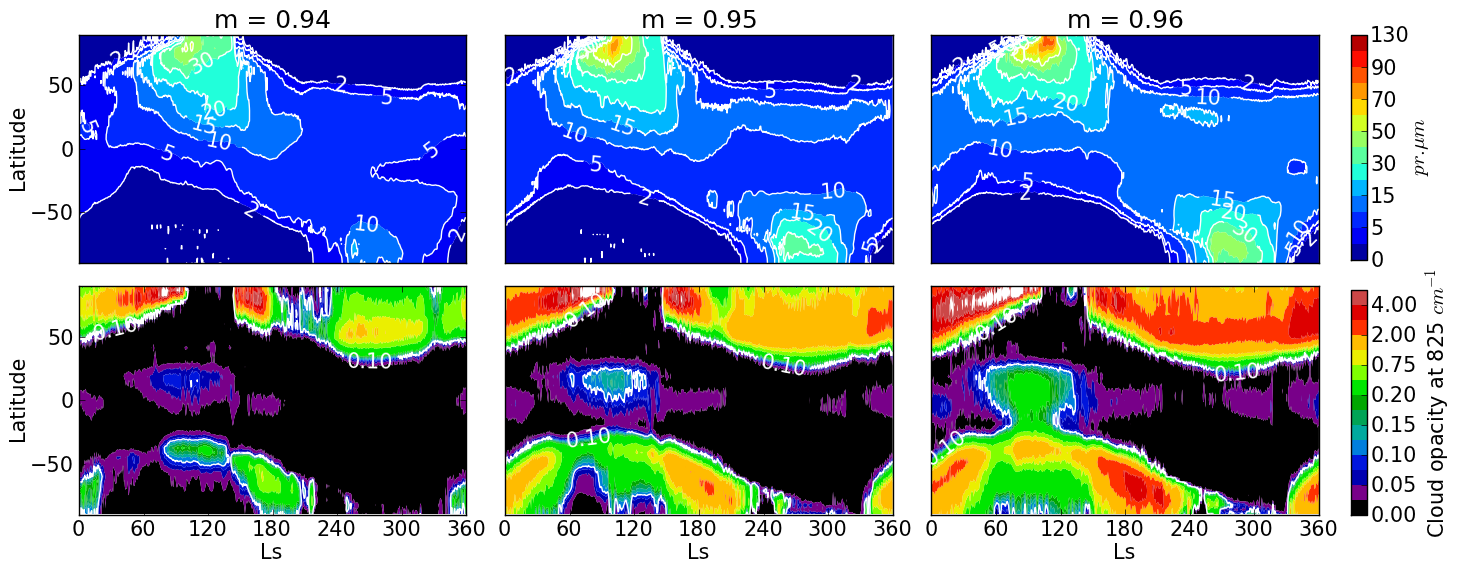}
\caption{GCM zonal mean quantities during daytime for MY26 of atmospheric water vapor (top row) and cloud opacity (bottom row) with three different values of contact parameter : 0.94 (left), 0.95 (middle) and 0.96 (right)}
\label{figP1}
\end{figure}

\begin{figure}[!ht]
\centering
\noindent\includegraphics[width=40pc]{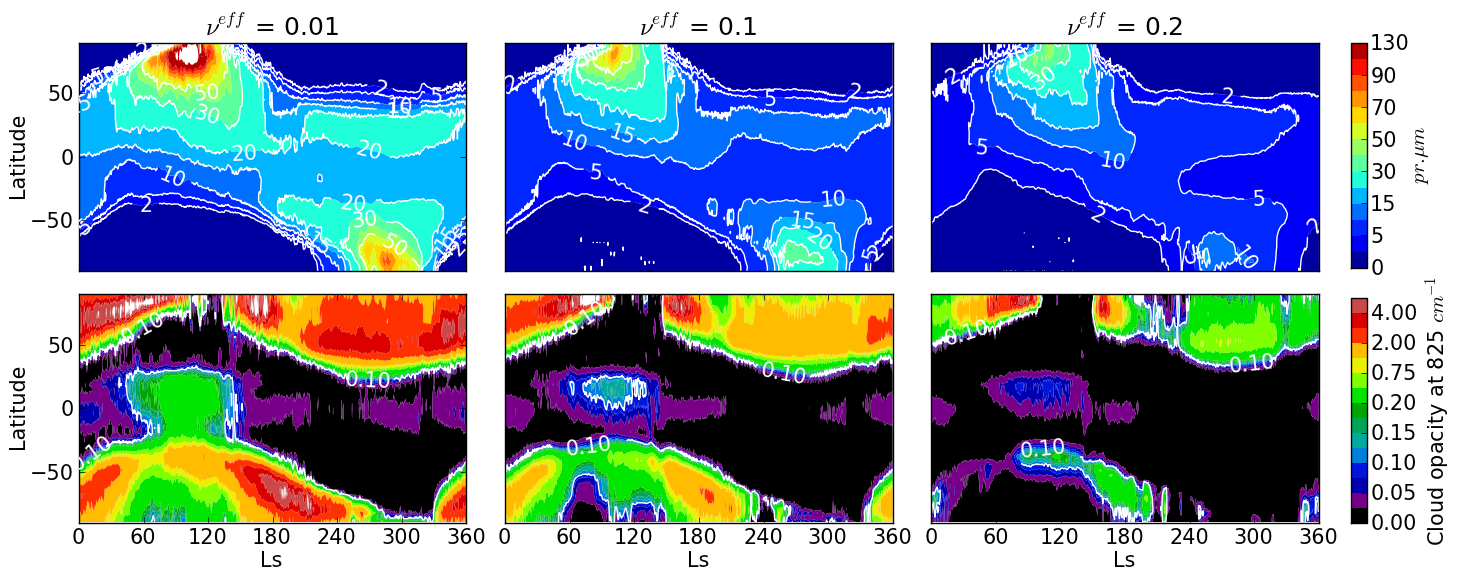}
\caption{GCM zonal mean quantities during daytime for MY26 of atmospheric water vapor (top row) and cloud opacity (bottom row) with three different values of cloud particle size distribution effective variance for sedimentation : 0.01 (left), 0.1 (middle) and 0.2 (right).
Note that we keep the value of the effective variance for the radiative transfer to the same value of 0.1 in all cases.
}
\label{figP3}
\end{figure}

The reference simulation is a basis for analysis and comprehension of the water cycle and of the sensitivity to the previously mentioned parameters.
Figures \ref{figP2}, \ref{figP4}, \ref{figP1} and \ref{figP3} explore the influence of parameters on the atmospheric water vapor and clouds after five years of simulation starting from the same reference steady state of the reference run.


A decrease of albedo or thermal inertia of the perennial surface ice increases the overall humidity and cloud opacity, especially in the aphelion cloud belt (Figures \ref{figP2} and \ref{figP4}).
With such parameters, quantities sublimated at the North Pole are larger than in the reference run due to warmer surface temperatures for the perennial surface ice.
Because of this, the period during which the North pole is free of clouds is reduced in these warm and wet cases, thus giving more weight to the undesired relation between clouds and the total water vapor released in the atmosphere, as explained in section \ref{globalmicrophy}.
The use of a smaller timestep for microphysics does not change the resulting cloud fields in simulations.
Thus, it remains unclear if this behaviour would be real in cases with higher surface temperatures or is a limitation of the model for these cases. 

Changing values of the ice thermal inertia or albedo has the same qualitative impact on the sublimation peak at the North pole, 
although the impact of each parameter on the diurnal cycle of surface temperatures of surface perennial ice is different.
A decrease of albedo or thermal inertia results in higher water vapor column values from Ls=250$^\circ$ to Ls=360$^\circ$ in the Northern hemisphere.
However, the higher sublimation peak near the South pole is more emphasized with the decrease of albedo than with the decrease of thermal inertia. 
This can be explained by the fact that thermal inertia only impacts the perennial surface ice on the North pole, whereas albedo impacts any surface ice thicker than 5 $\mu$m. 
Hence, seasonal frost sublimation is enhanced in those two places when the ice albedo is decreased.
Conversely, an increase of albedo, or thermal inertia, tends to dry out the water cycle and produce fainter aphelion clouds.

\subsubsection{Nucleation contact parameter and effective variance of the cloud particle size distribution}

The contact parameter controls the amount of activated nuclei in water ice clouds.
Increasing the contact parameter, i.e. the efficiency of nucleation, increases the number of ice particles for a given cloud mass (as long as all CCN are not activated).
Hence, the size of ice particles is lowered and the sedimentation of clouds is less effective.
This impacts the transport of water from polar regions to lower latitude regions, which is made easier due to longer lasting clouds \citep{Mont:04}.
It follows on that the atmosphere keeps more water vapor (figure \ref{figP1}).
The same statement stands for the effective variance of sedimentation (figure \ref{figP3}): a decrease of the variance decreases sedimentation, and increases the global amount of water vapor in the atmosphere.
Varying the contact parameter between 0.94 and 0.96 as presented here changes the number of ice particles by two orders of magnitude.
This helps to prevent the formation of polar clouds above the North pole around summer solstice, more efficiently than what a change of the effective variance for sedimentation would produce.
This is shown in the left panels of figure \ref{figP1}, where the sublimation peak at the North pole is less pronounced than the three other wet cases for the other parameters.
In other words, increasing the contact parameter tends to increase the water vapor column in the tropics rather than the vapor near the North pole.
Conversely, an increase of sedimentation due to a smaller contact parameter, or a larger effective variance, limits the transport of water, and tends to dry the atmosphere. 

\section{Analysis and Comparison with Observations} \label{secdiscuss}

\subsection{Effect of Radiatively Active Clouds on the Global Water Cycle}\label{RACanalysis}
\begin{figure}
\centering
\noindent\includegraphics[width=40pc]{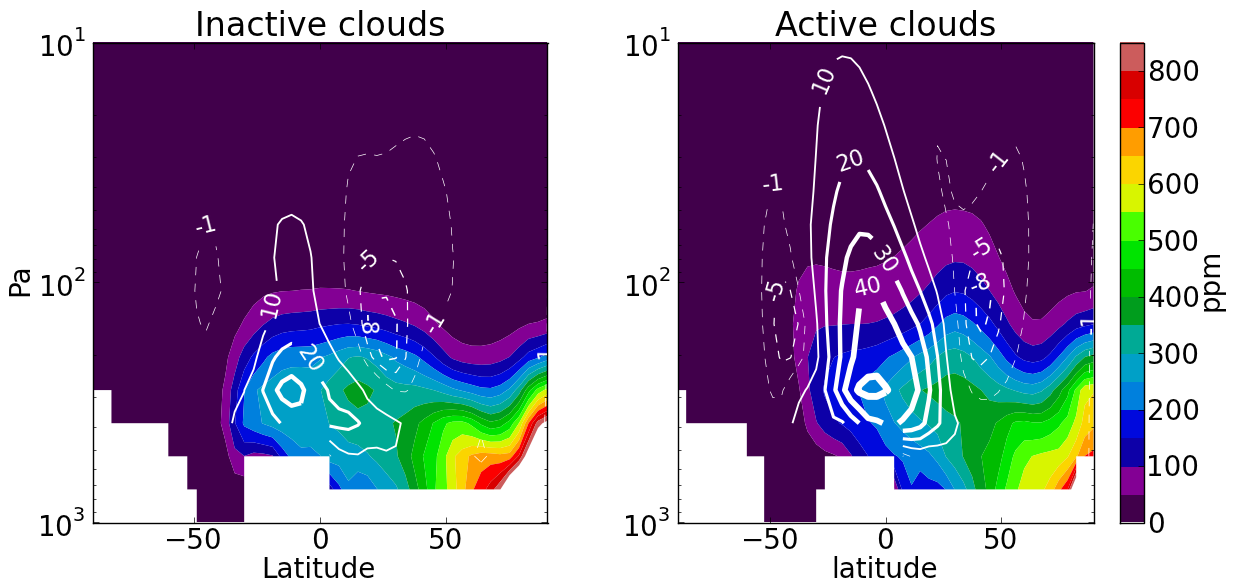}
\caption{Zonal mean of atmospheric water vapor (colours, volume mixing ratio), and stream function (10$^8$ kg/s contoured
lines) averaged over Ls=120$^\circ$-150$^\circ$ with radiatively active clouds (right) and without (left)}
\label{figC}
\end{figure}

One simple and obvious way to assess the effect of radiatively active clouds on the modeled global water cycle is to compare simulations with and without radiatively active clouds.
All other things being equal, the main difference is the amount of vapor sublimated at the North pole that is almost twice lower with radiatively inactive clouds than with active clouds (not shown).
This discrepancy is caused both by the direct effect of radiatively active clouds on the surface temperature before Ls=90$^\circ$, and the induced stronger baroclinic wave activity at the same period that generates stronger winds and enhances sublimation flux from surface to atmosphere, as can be inferred from section \ref{sublimationscheme}. 
The global amount of water vapor is primarily controlled by this sublimation peak at the North pole.
To explore other factors than this dominant one, we lowered the surface ice albedo and thermal inertia in the simulation with radiatively inactive clouds to increase the surface temperature and thus the sublimating flux of vapor.
The most prominent difference between a simulation with radiatively active clouds and one without is then the amount of atmospheric water vapor that is advected from the North pole to the tropics during northern spring and summer, as seen in figure \ref{figC}.

This is due to the strengthening of the Hadley cell with radiatively active clouds \citep{Made:12radclouds} which traps more efficiently water vapor at latitudes 40$^\circ$ to 60$^\circ$ N.
Indeed, the vertical extension of the Hadley cell above the hygropause is more emphasized with radiatively active clouds than without.
This manifestation of the so-called ``Clancy effect'' \cite{Clan:96b} is reinforced by radiatively active clouds and changes the equatorward transport of water vapor. 


\subsection{Effect on Dust Vertical Distribution}
\begin{figure}
\centering
\noindent\includegraphics[width=20pc]{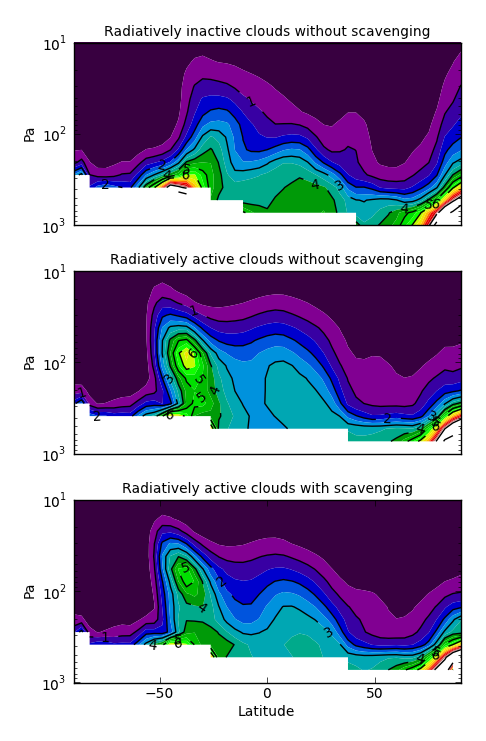}
\caption{Mass mixing ratio of airborne dust (ppm) at longitude 25$^\circ$W during the period Ls=90$^\circ$-120$^\circ$ for a GCM simulation without radiatively active clouds (upper panel), with radiatively active clouds but without scavenging (middle panel), with both radiatively active clouds and scavenging (lower panel).}
\label{figK}
\end{figure}

One motivation for the inclusion of dust scavenging by water ice clouds was to assess its influence on dust vertical distribution.
In particular, it has been suggested that scavenging could explain the formation of the detached dust layers observed by MCS \citep{McCl:10,Heav:11mcs} by the enrichment of dust below the ice clouds (by releasing the dust when the cloud sublimates) or by depletion in the clouds (by decrease of dust quantities due to nucleation within the cloud).
It is possible to assess the effect of scavenging on the whole simulation, and especially on the dust distribution by simply turning off the interactions of clouds with airborne dust in the simulations.
In other words, if a cloud forms, CCN are activated but no dust is removed from the ambient atmosphere.
When ice particles sublimate, the released CCN just disappear and are not turned back into dust.
From the reference simulation, it appears that turning on and off scavenging does not help to produce more dust detached layers.
However, even without scavenging, the radiative effect of clouds sometimes favors the formation of such layers, as illustrated in the example of figure \ref{figK}.

This is due to the reinforcement of semi-diurnal waves with radiatively active clouds \citep{Hins:04clouds} that locally enhance upward winds.
However, such layers only occur at rare and specific locations in our simulations.
The example given in figure \ref{figK} is the most evident one with a detached layer, and is not representative of the dust vertical distribution throughout the simulation.
Therefore radiatively active clouds, alone, cannot account for the formation of detached layers of dust, but it is worth noting their possible contribution in combination with other possible mechanisms \citep{Spig:13rocket,Rafk:12,Heav:11dust}.

\subsection{Supersaturation}\label{supersat}
The use of our new microphysical scheme for the formation of water ice clouds opens the possibility to simulate supersaturation.

Above the hygropause, the lack of dust particles to trigger nucleation allows water vapor to remain in a supersaturated state.
In that case, the most prominent process that affects water vapor is the photodissociation of water vapor.
Because of the exponential relation of saturation with temperature, a few ppm of water vapor can correspond to a huge saturation ratio, and photodissociation cannot be neglected.
Therefore, we used the photochemistry module implemented in the GCM \citep{Lefe:04} to study the supersaturation of water vapor in dedicated simulations.
The model is vertically extended when using this module to take into account all photochemical processes that take place at different altitudes. 
The use of the photochemistry module and this vertical extension do not significantly change the simulated water cycle, as quantities of supersaturated water vapor affected by the photochemistry are very small when compared to the typical amount of water vapor (less than 2 ppm vs. hundreds of ppm).
The photochemistry module and this extension were not used for previous simulations, as it is computationally expensive.

\begin{figure}
\centering
\noindent\includegraphics[width=40pc]{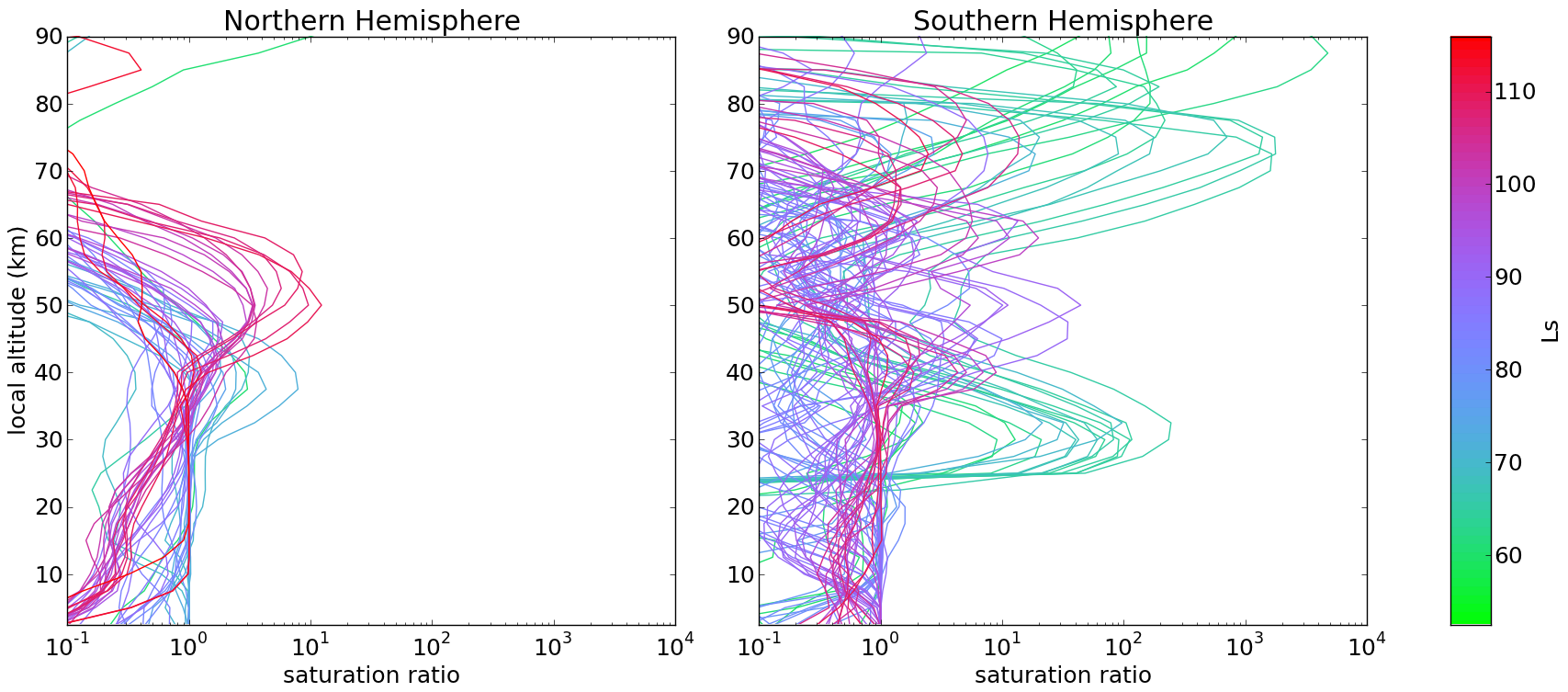}
\caption{Water vapor saturation profiles at the same locations than observed by SPICAM in \cite{Malt:11}}
\label{figI}
\end{figure}

Figure \ref{figI} shows supersaturation profiles from the GCM, at the same locations as those reported from SPICAM solar occultation measurements in \citet{Malt:11}.
When compared to figure 3 of \citet{Malt:11}, the Northern hemisphere values are in agreement with the observations, with a typical saturation ratio S of at most 10 between 30 and 50 km before Ls=90$^\circ$ and between 40 and 60 km after.
In the Southern hemisphere, some profiles after Ls=75$^\circ$ match the observations, but about 10 km higher than observed.
Moreover, there is a clear discrepancy before Ls=75$^\circ$ with peaks at S=100 between 20 and 40 km, and S$>$1000 above 60 km.
This can be explained by the vicinity of those profiles with the strong temperature gradient of the polar night.

\begin{figure}
\centering
\noindent\includegraphics[width=40pc]{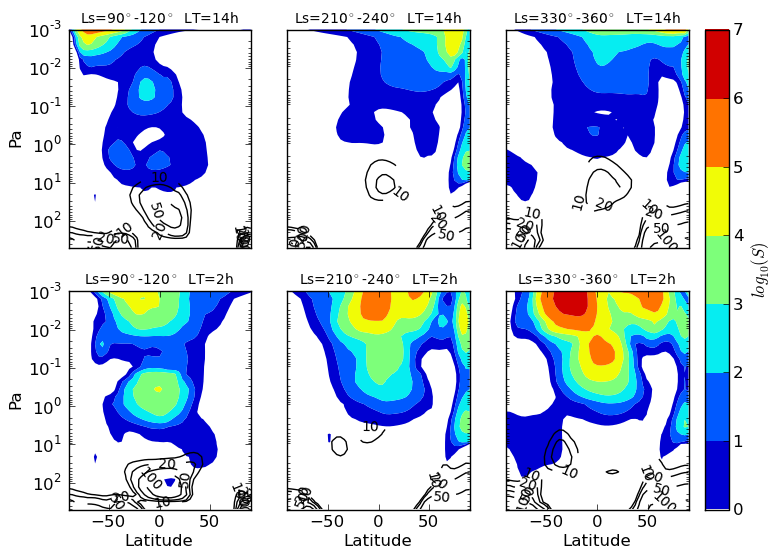}
\caption{Base-10 logarithm of water vapor saturation ratio (shaded) and water ice volume mixing ratio in ppm at local times 14h (upper panels) and 2h (lower panels), averaged over Ls=90$^\circ$-120$^\circ$ (left), Ls=210$^\circ$-240$^\circ$ (middle) and Ls=330$^\circ$-360$^\circ$ (right) }
\label{figI2}
\end{figure}

Figure \ref{figI2} shows the structure of supersaturation up to 0.001 Pa at three different seasons, as predicted by the GCM. Supersaturated water vapor remains above the water ice clouds.
The pattern of supersaturation varies significantly with local time, mainly because of the diurnal cycle of temperature.
At Ls=210$^\circ$-240$^\circ$ and Ls=330$^\circ$-360$^\circ$, temperature clearly controls the boundary of supersatured water vapor.

The GCM fails to form dust detached layers, and might also miss the actual distribution of dust above the hygropause \citep{Heav:11mcs,Maat:13} that coexists with supersatured water vapor \citep{Malt:13}.
Therefore, we can wonder whether a supersatured ratio above $10^3$ is realistic or not.
It reaches the limits of our knowledge of dust distribution at these altitudes.
Even without the presence of dust, homogeneous nucleation (the formation of ice without another support) could be taken into account for such saturation ratios, which is not the case in the present model and is significative for a ratio S$>$10$^5$ at 50 km \citep{Maat:05}.
All in all, such considerations do not affect the total amount of atmospheric water in simulations because this supersatured water vapor corresponds to negligible amounts of water vapor, less than one ppm.

\subsection{Polar Ground Ice Budget}
\begin{figure}
\centering
\noindent\includegraphics[width=40pc]{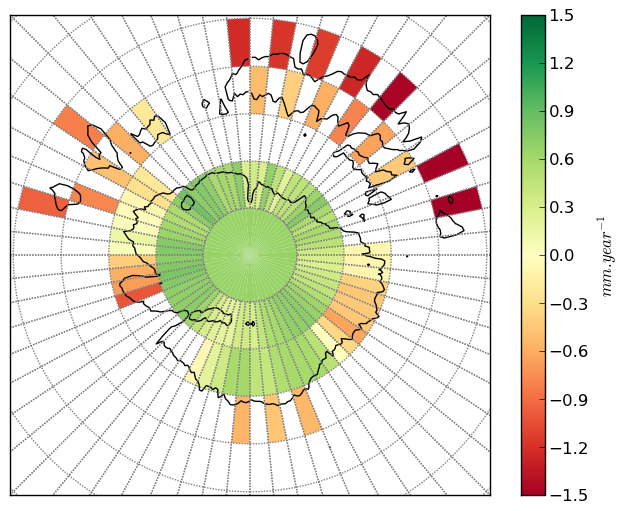}
\caption{
Annual budget of permanent surface ice at Ls=0$^\circ$ for the perennial surface ice on North pole as defined in the GCM in mm/year. White indicates this mesh grid is not defined as a permanent ice reservoir in the GCM. Positive values indicate an increase of ice. A contour of albedo derived from TES observations is shown to indicate the distribution of ice on a finer scale.}
\label{figB}
\end{figure}

The improved representation of permanent surface ice on the North pole gives insights on the fate of polar ice.
Figure \ref{figB} shows that the long term stability of permanent ice depends on the latitude of the considered deposit.

Ice is accumulating on the central cap at latitudes above 80$^\circ$N, whereas ice outliers are being hollowed at a rate of 1 mm per year.
Conversely, in southern polar regions, ice is accumulating at a rate of 0.5 mm per year (not represented).
This instability of surface ice that does not significantly vary on a few years timescale, and therefore supposed as perennial in atmospheric models, could be of great interest for recent climate changes and paleoclimatology.
However, given the current uncertainties and limitations of the model to represent the water cycle, this question requires further investigation and is out of the scope of the present paper.
The current version of the GCM suggests that the water ice reservoirs are not in an annual equilibrium state. 

\subsection{Exploration of Parameters Space and Tuning of the Water Cycle} \label{spaceparams}

The issue about the dry tropics raised in section \ref{refrun} and explained in \ref{RACanalysis} motivates a change of the values of these model parameters towards wetter simulations.
A change of albedo, thermal inertia or effective variance would indeed increase the water vapor at the tropics, but the maximum of water vapor at the North pole would also be increased as seen in section \ref{sensitivityparams}, in stark contrast with observations.

\begin{figure}
\centering
\noindent\includegraphics[width=40pc]{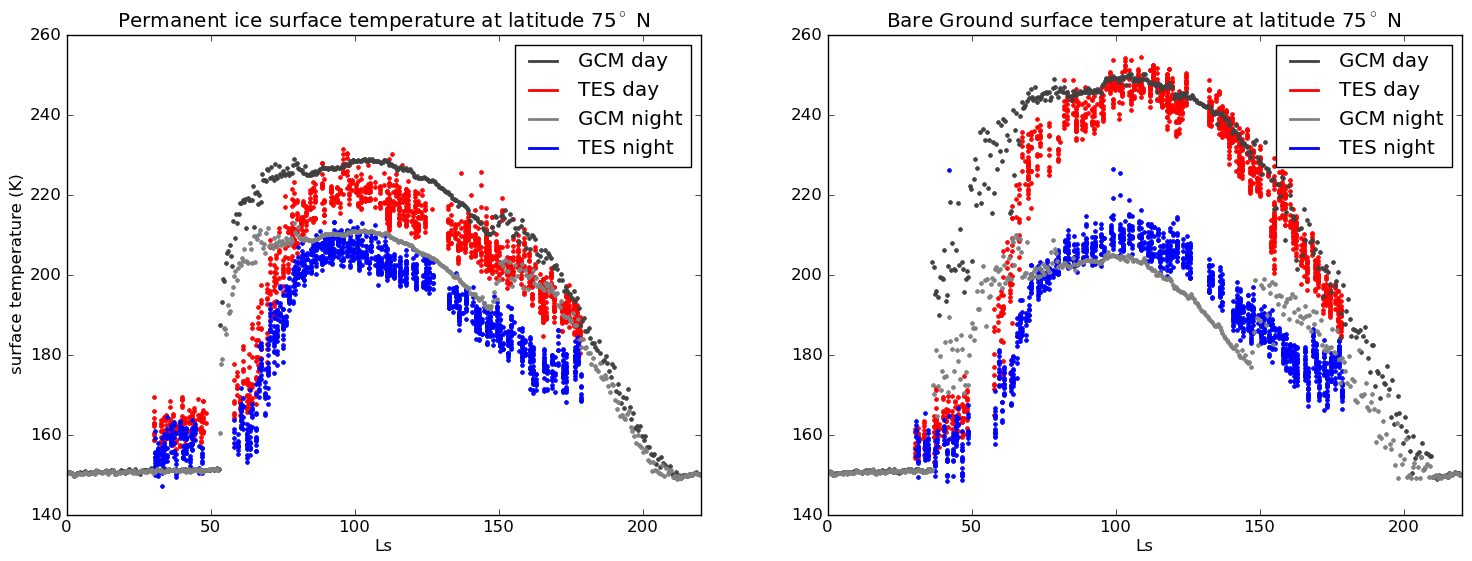}
\caption{Surface temperature of GCM and TES at 75$^\circ$N at local times 14h and 2h for two patches of permanent surface ice (left) and bare ground (right)}
\label{figH}
\end{figure}

Furthermore, a decrease of albedo or thermal inertia would result in greater surface temperatures than in the reference simulation.
However, this would not be realistic because modeled temperatures are already slightly too high compared to the observations, as shown in figure \ref{figH}.

One can see that the values of albedo and thermal inertia selected for the reference run yields temperatures 5 K warmer on the average than observed for permanent surface ice.

\begin{figure}
\centering
\noindent\includegraphics[width=20pc]{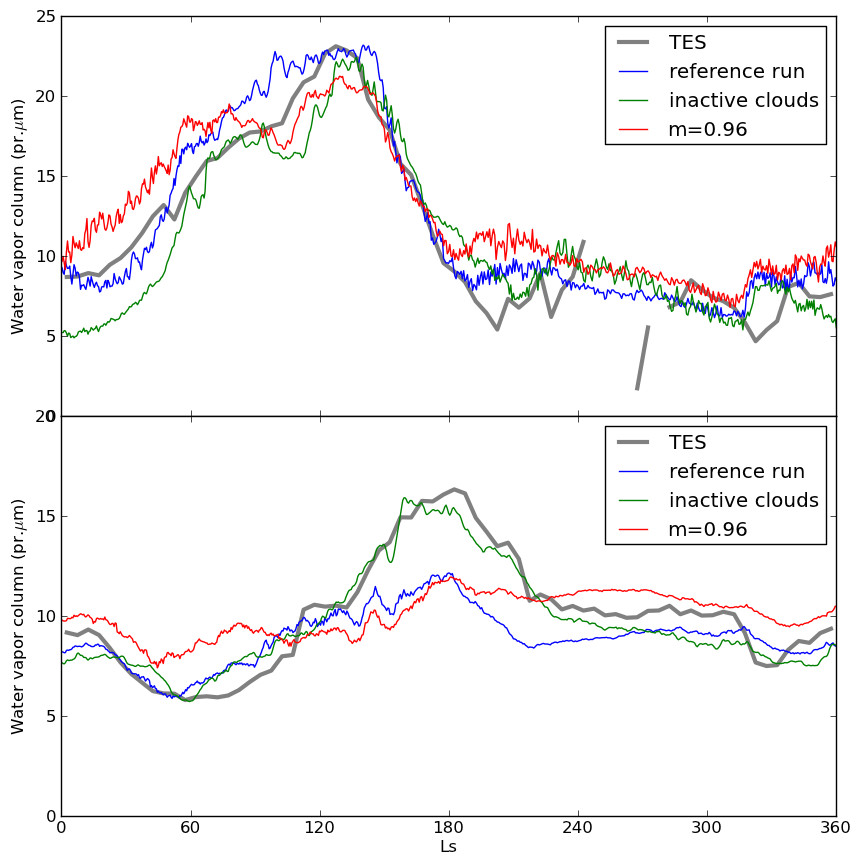}
\caption{Zonal mean quantities during daytime for MY26 of atmospheric water vapor at latitude
40$^\circ$N (top) and equator (bottom) of TES (grey) and GCM for three different runs : reference run
(blue), same as reference run but with contact angle m = 0.96 (red) and same as reference run but without radiatively active clouds and different tuning (see text) (green).}
\label{figAbis}
\end{figure}

Another possibility would be to change the distribution of perennial surface water ice, described in section \ref{permice1}.
The extent of this distribution, shown in figure \ref{figD}, directly controls the amount of water vapor released by surface sublimation in polar regions.
However, as mentioned in Section~\ref{permice1} the amount of perennial surface ice is already slightly greater in the GCM at latitude 75$^\circ$N than in the 
surface ice distribution retrieved from TES temperature observations (see the lower right panel of figure \ref{figD}). 
Thus, further increasing the coverage of perennial surface ice would not be realistic. 
An attractive option would be to increase the value of the contact parameter $m$ to increase the amount of water vapor in the tropics.
Figure \ref{figAbis} shows that increasing the value of the contact parameter helps to match the quantities of water vapor at the equator past Ls=180$^\circ$.
However, the maximum of water vapor around Ls=180$^\circ$ is then less well represented.

The increase of water vapor in northern latitudes between Ls=120$^\circ$ and Ls=180$^\circ$ is controlled by the advection of water vapor sublimated at the North pole, whereas the decrease after Ls=180$^\circ$ is controlled by the sedimentation flux of ice particles in the tropics.
Increasing $m$ affects the size of particles and therefore modifies the water vapor decrease after Ls=180$^\circ$.
It has no effect on the lack of southward vapor advection due to a stronger Hadley cell.
A representation of the water vapor closer to observations at equator could only be obtained without radiatively active clouds, which yields a correct maximum at 15 pr.$\mu$m at Ls=180$^\circ$, even though the maximum is well represented at 40$^\circ$N at Ls=120$^\circ$ for the three cases (inactive clouds, active clouds with m=0.95, and m=0.96).

This raises the question of the improvement of the water cycle with the current version of the GCM.
Even if it is possible to assess the local effect of each parameter, the complete simulation is a complex balance after at least a decade of integrations before reaching a steady state, and the best way to quantitatively assess the global impact of one single parameter is through an empirical approach.
In such a system, effects of parameters are coupled and no better set of parameters than the one used for the reference run has been found. 
Given the current GCM configuration, we cannot find any significant leverage on the strength of the Hadley cell after summer solstice without degrading the water cycle.

\section{Conclusion}

We have included a new microphysical scheme in a Global Climate Model that takes into account the radiative effects of clouds to represent cloud particles nucleation on dust particles, ice particle growth, and scavenging of dust particles resulting from the condensation of ice.
The use of this microphysical scheme prevents the formation of clouds above the poles around summer solstice that occurs when microphysics is not well taken into account, and which has been found to negatively alter the modelling of the water cycle in many Martian GCMs taking into account the radiative effect of clouds (\citealt{Made:12radclouds,Urat:13h2o,Habe:11} and R. J. Wilson, personal communication).
This enables a more realistic representation of the water cycle with radiatively active clouds.
For this purpose, we have explored the model sensitivity to model parameters not well known from observations in order to tune the water cycle to match available observations and in particular TES data.

This study leads us to conclude that the radiative effects of clouds are (at least) twofold:
First, there is an impact on the cloud formation itself, due to the local change in atmospheric temperature that generates a feedback on the cloud formation.
Handling the cloud formation with such radiative effects is critical for the simulation of the Martian water cycle with a GCM.
Second, there is an impact on the Hadley circulation that is reinforced in comparison with simulations without radiative effects of clouds.
This changes the equatorward transport of water, leading to a dry bias in tropics when compared to observations.
The exploration of the influence of the four physical parameters on the water cycle leads us to conclude that this dry bias cannot be completely 
corrected with the current representation of the water cycle in the GCM.
Nevertheless, overall the new model is more consistent with observations than before, thanks 
a simulated thermal structure that is more consistent with observations because of the use of radiatively active clouds which lead to
a qualitative improvement of the modeled water cycle. The coupling between atmospheric temperatures and the water cycle, through clouds and global circulation is so strong
that the modeling of these two coupled quantities should be seen as an inseparable whole.
\\
This new representation also allows the exploration of new topics, such as vapor supersaturation in the Martian atmosphere, the long-term stability of perennial surface ice, and the question of missing sources and sinks in the water cycle.
It has also been found that dust scavenging by water ice clouds are most probably not the cause of dust detached layers observed in the Martian atmosphere, even though atmospheric waves enhanced by the radiative effect of clouds sometimes help the formation of detached layers.
The inclusion of new physical processes allowed us to point out model sensitivities that should be addressed before adapting any GCM to paleoclimates, where prevailed warmer and wetter conditions than today and for which radiative effects of clouds has long been neglected in GCM studies.

\section*{Acknowledgements}
\thispagestyle{empty}

We thank the financial support provided by the Centre National d'\'Etudes Spatiales (CNES) and the Observatoire de Paris for the Labex Exploration Spatiale des Environnements Plan\'etaires (ESEP) (N$^\circ$ 2011-LABX-030), through the ANR ``Investissements d'avenir'' via the ``Initiative d'excellence'' PSL* (convention N$^\circ$ ANR-10-IDEX-0001-02).
The development of the LMD GCM is supported by CNES, Centre National de la Recherche Scientifique (CNRS), and European Space Agency (ESA), in collaboration with Instituto de Astrof\'isica de Andaluc\'ia (IAA, Granada), Open University, and Atmospheric, Oceanic, and Planetary Physics laboratory (AOPP, Oxford University).
We also wish to thank Luca Montabone and M.D. Smith for helping with the TES data, Luca Maltagliati for providing the SPICAM vertical profiles of supersaturation, and Franck Lef\`evre for his useful feedbacks.
\clearpage

\bibliographystyle{apalike}
\bibliography{../newfred}

\end{document}